\documentstyle[12pt,epsf,epsfig,wrapfig]{article}

\newcommand{\rar}{\rightarrow}
\newcommand{\pdup}{p_\uparrow}

\newcommand{\pplup}{\pdup + p \rar \pi^+ + X} 
\newcommand{\pimp}{\pi^- + \pdup \rar \pi^0 + X}

\newcommand{\xf}{x_{\mathrm F}}
\newcommand{\pdupp}{\pdup + p \rar \pi^0 + X}

\begin{document}
\begin{center}
{\bfseries PRESENCE OF THE UNIVERSAL SUBSTRUCTURES IN THE  
HADRONS -- CONSTITUENT QUARKS}
\vskip 5mm

{\underline {V.V. Mochalov}, S.M. Troshin$^{\dag}$} and
A.N. Vasiliev
\vskip 5mm
{\small {\it Institute of High Energy Physics, Protvino, Russia}

$\dag$ {\it
E-mail: troshin@mx.ihep.su}}
\end{center}

\begin{abstract}
The universality of single-spin asymmetry on inclusive $\pi$-meson 
production is discussed. This universality can be related to
the hadron substructure --- constituent quarks in the frame of the quark model
for U-matrix.
\end{abstract}

\vskip 5mm

Polarization experiments give us an unique opportunity to probe the 
nucleon internal structure. While spin averaged cross-sections can be
calculated within acceptable accuracy, current theory of strong 
interactions can not describe large spin asymmetries and polarization. 
Polarization is a precise tool for measuring the electroweak 
parameters, spin dependent structure functions {\it etc}. 
After establishing the fact that the nucleon spin
is not described by simple summing of the quark spins, the
study of gluonic and orbital momentum contribution
to it is very important and intriguing. 

\begin{wrapfigure}{R}{6.2cm}
\mbox{\epsfig{figure=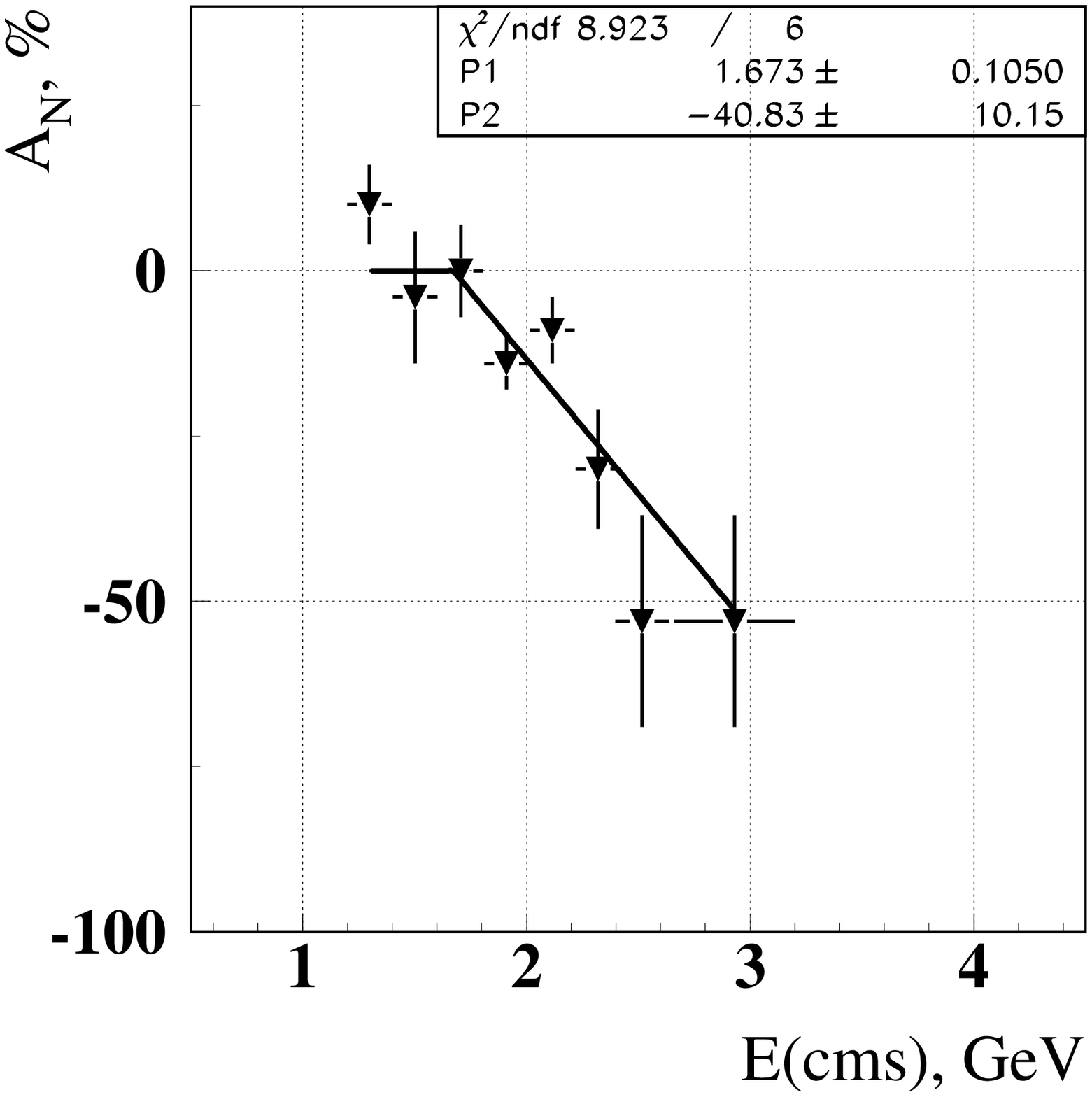,width=5.8cm,height=5.2cm}}
\mbox{\epsfig{figure=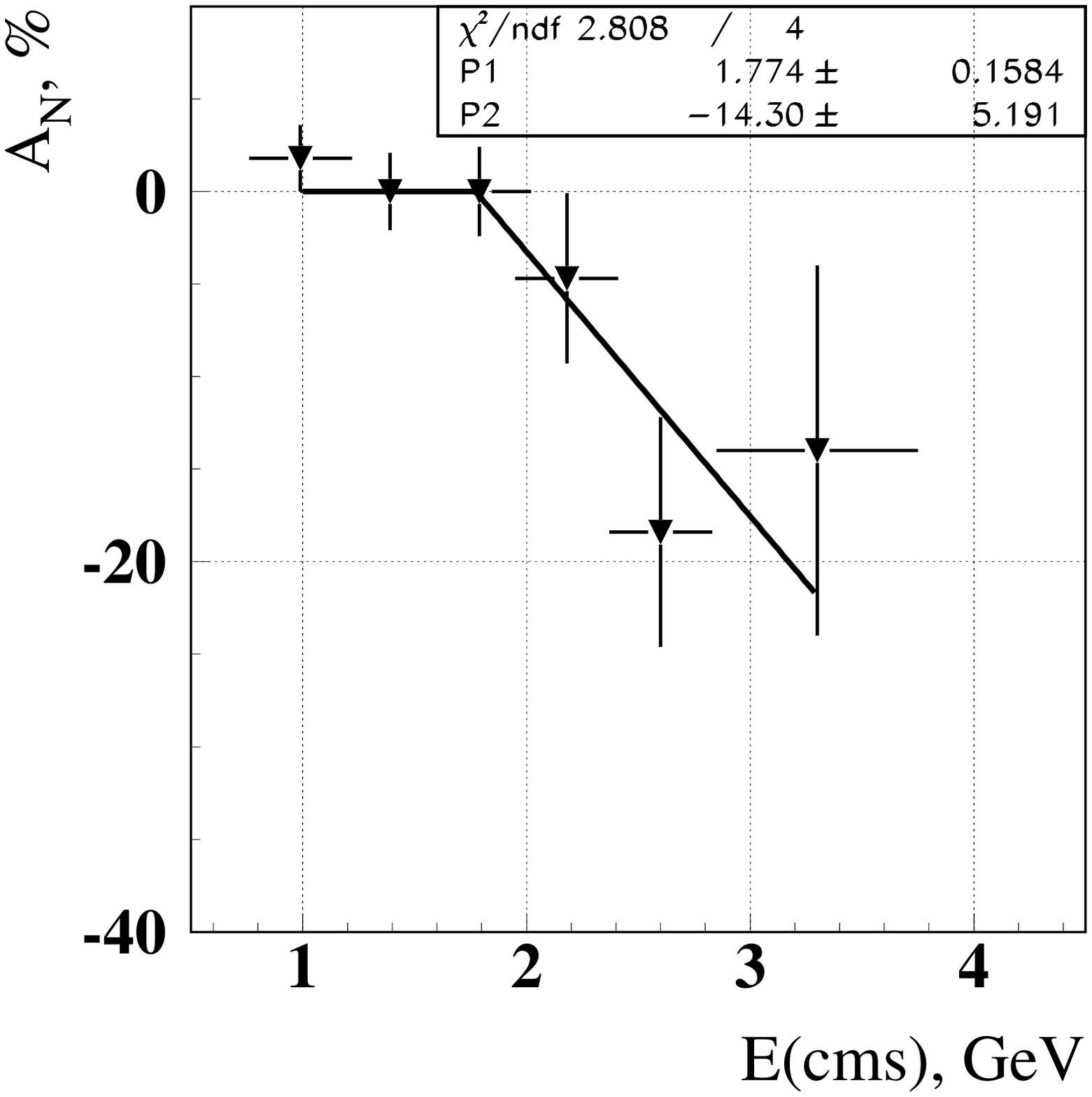,width=5.8cm,height=5.2cm}}\\
{\small {\bf Figure 1.}  $A_N^{\pi^0}$  in the reaction $\pimp$ 
at the central region (top, \cite{protv88}) and in the  target 
fragmentation region (bottom,  \cite{proza40}) at 40~GeV.
}
\end{wrapfigure}

Unexpected large values of  single spin asymmetry $A_N$ (SSA)
in inclusive $\pi$-meson production are real challenge
to current theory because perturbative Quantum Chromodynamics 
predicts small asymmetries decreasing with transverse momentum. 
Various models were developed to explain results from E704 (FNAL), 
PROZA-M and FODS (both Protvino) and several BNL experiments. 
Most of the models analyse experimental data in terms of $\xf$ and/or $p_T$.  
To investigate the dependence of SSA on a secondary meson production
angle, the measurements  in the reaction $\pimp$ 
were carried out at the PROZA-M experiment (Protvino) at 40~GeV pion 
beam in the two different kinematic regions:
at Feinman scaling variable $\xf \approx 0$ \cite{protv88} and 
in the polarized target fragmentation region \cite{proza40}.  
There is an indication that the asymmeties are zero till some 
threshold value and then begin to grow up linearly.  
In this case we can fit SSA by the function 

$A_N = \left\{ \begin{array}{ll} 0 & \textrm {, если $E<E_0$} \\
k \cdot (E-E_0) & \textrm {, если $E \geq E_0$} 
\end{array} \right.$

\noindent with two parameters -- threshold energy $E_0$ 
and $k$. 

  It was reported at this Workshop \cite{proza40} that the asymmetry of 
inclusive $\pi^0$ production in the reaction $\pimp$  begins to grow up 
at the same centre of mass energy $E_0^{cms} \approx 1.7$~GeV 
(See  {\bf Fig.~1}).

\begin{wrapfigure}{R}{10.4cm}
\mbox{\epsfig{figure=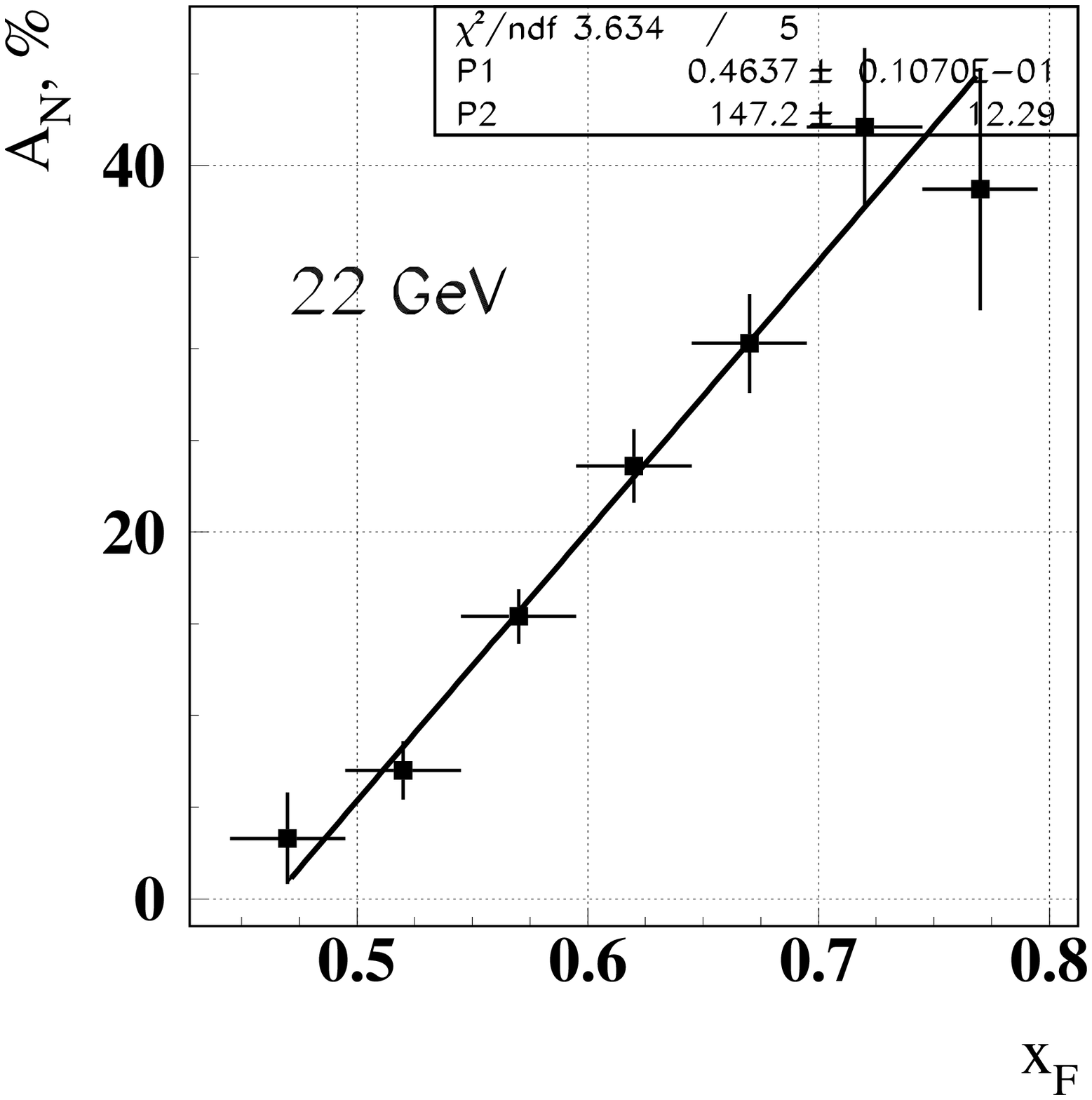,width=5.cm,height=5.cm}}
\mbox{\epsfig{figure=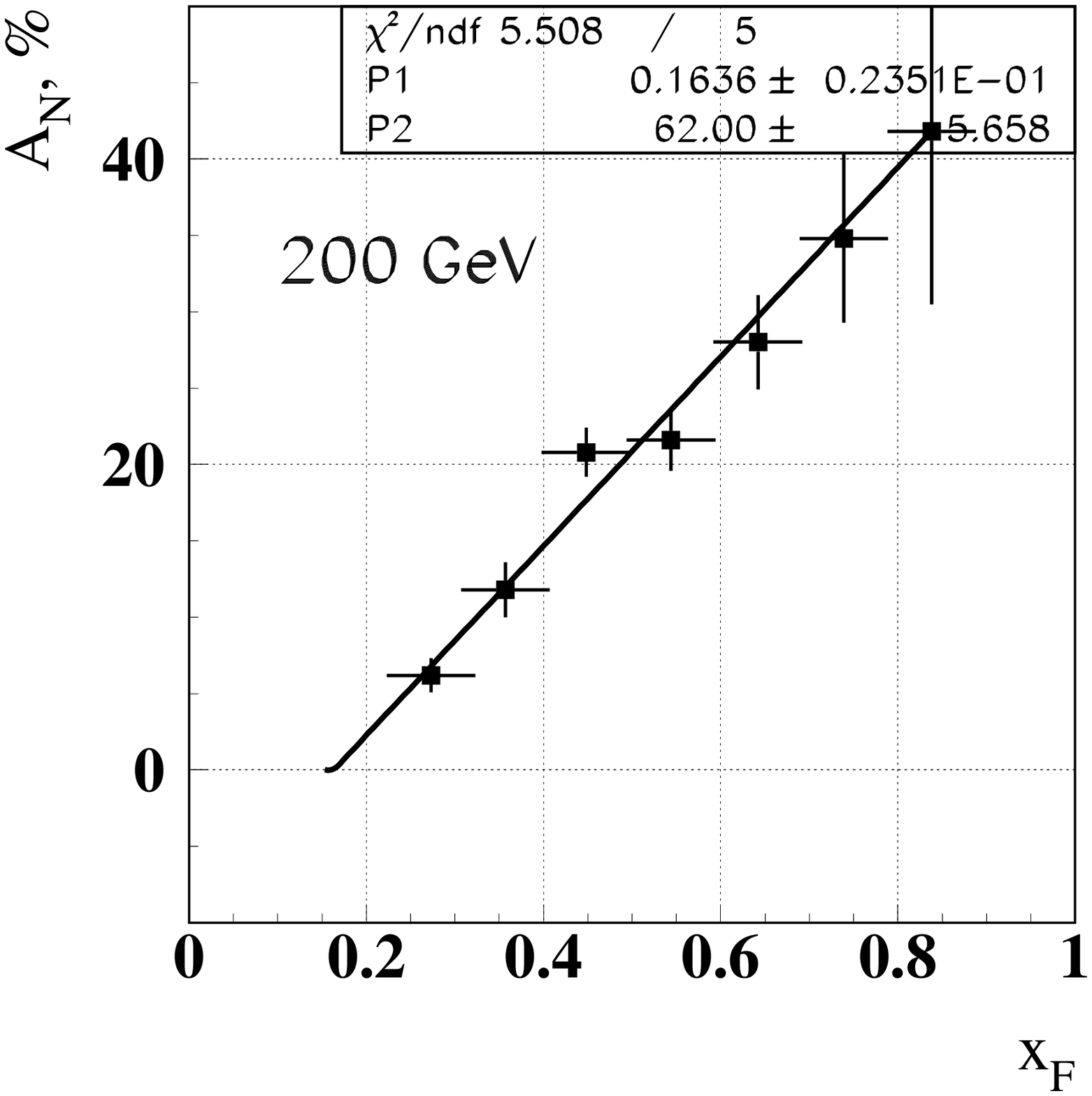,width=5.cm,height=5.cm}}\\
\mbox{\epsfig{figure=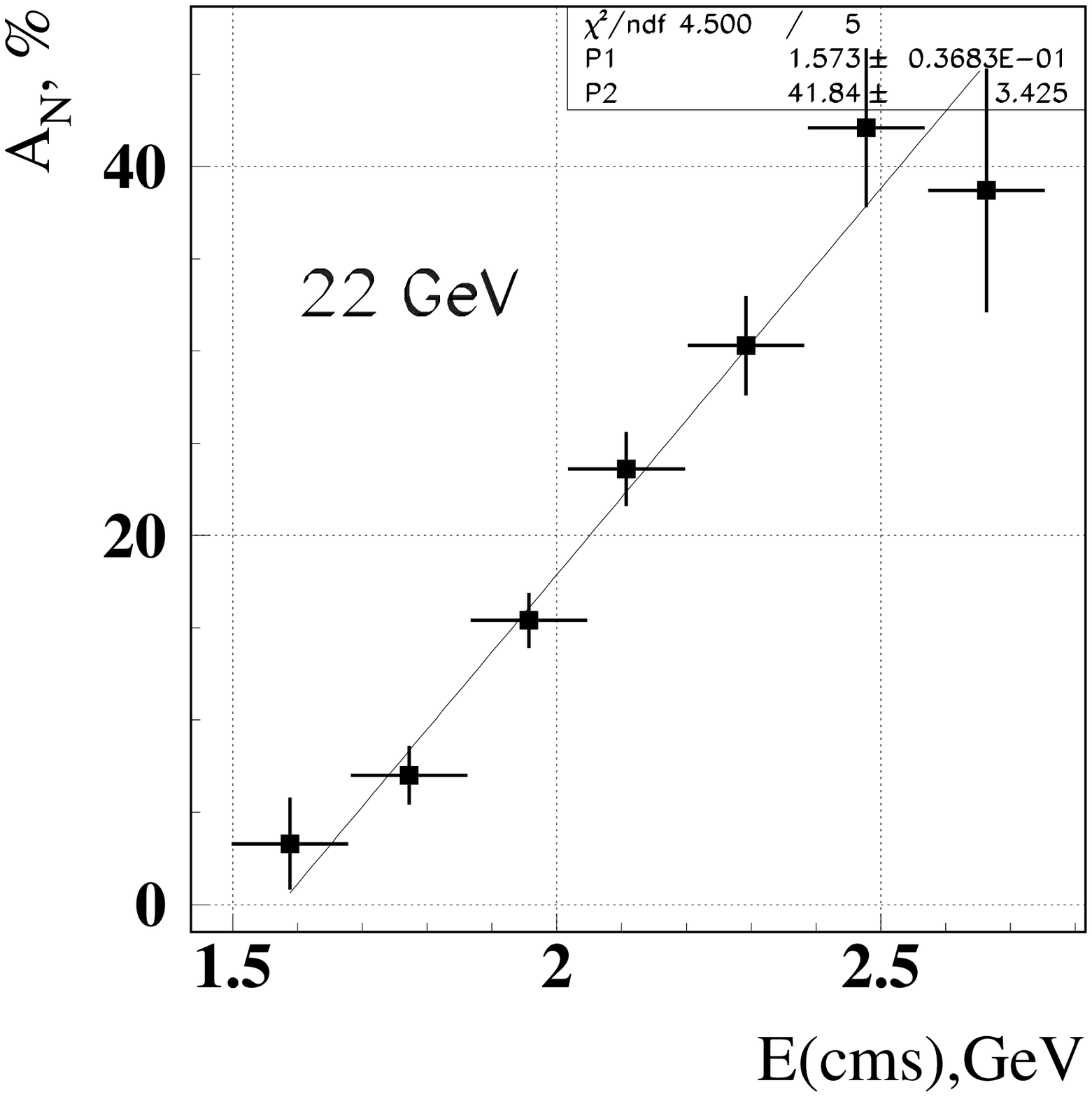,width=5.cm,height=5.cm}}
\mbox{\epsfig{figure=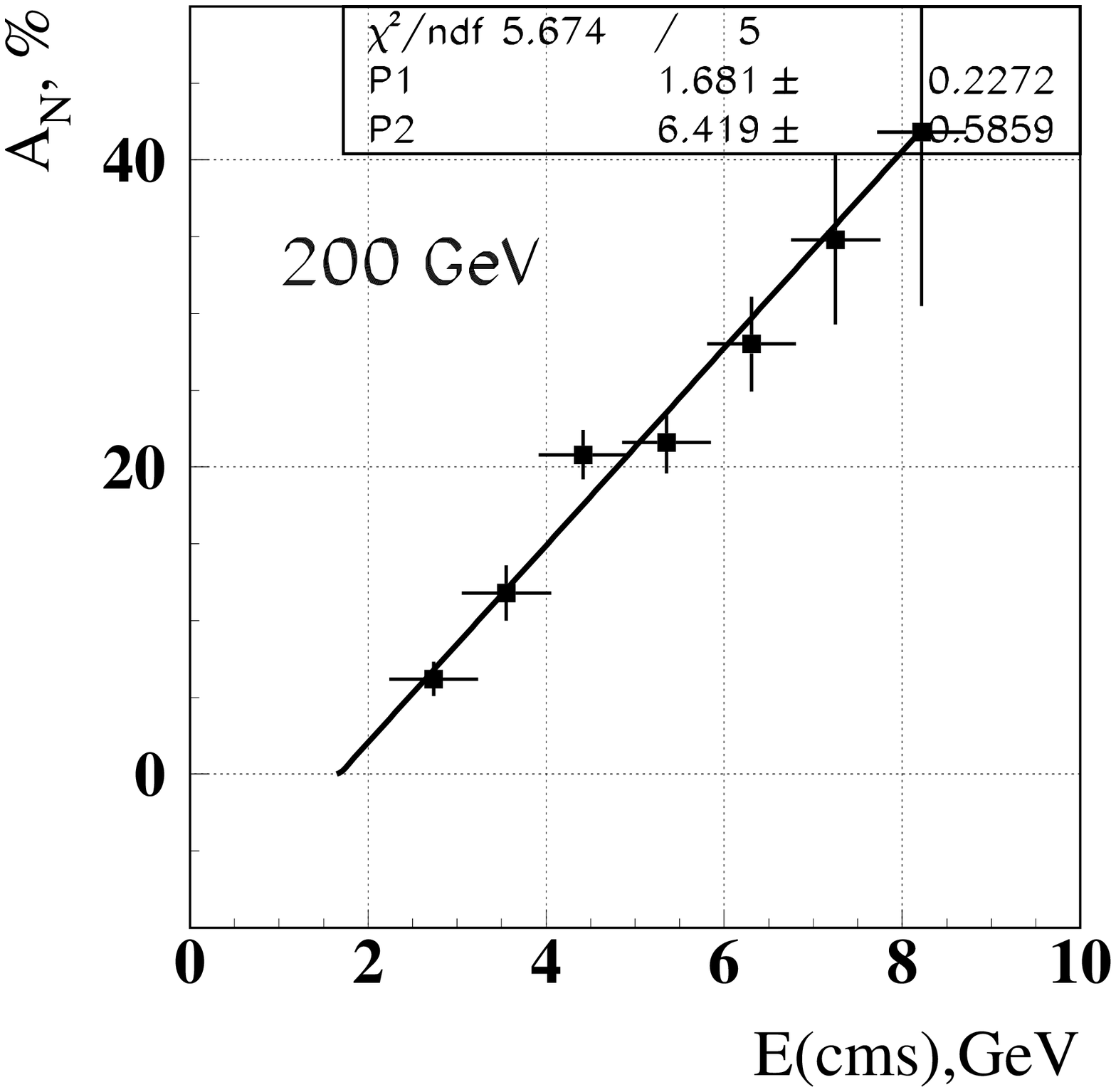,width=5.cm,height=5.cm}}\\
{\small {\bf Figure 2.}  The  dependence of $\pi^+$ asymmetry on $\xf$ (top) 
and energy (bottom) in the reaction $\pplup$ 
at 22~GeV (left, \cite{e925}) and 200~GeV (right, \cite{e704fragm}).
}
\end{wrapfigure}

Nevertheless from this statement we can not make the 
final conclusion whether the SSA behaviour 
depends on a beam energy or not. We analysed other experimental data 
to study this threshold effect. 
A comparison of E704 (FNAL) and E925 (BNL) experimental results is 
presented in {\bf Fig.~2}.

The $\pi^+$ asymmetry in E925 experiment (22~GeV, \cite{e925}) 
and in  E704 experiment (200~GeV, \cite{e704fragm})    
begins to rise up at different values of $\xf$ 
($\xf^0 \approx 0.18$ for E704 and $\xf^0 \approx 0.46$ for E925). 
It was also found that the asymmetry for these two experiments 
begins to grow up at the same longitudinal or full energy in the centre 
of mass system, $E_0^{cms} \approx 1.6$~GeV. It happened to be 
surprisingly the same energy as for the PROZA-M experiment.    

\begin{wrapfigure}{R}{10.cm}
\mbox{\epsfig{figure=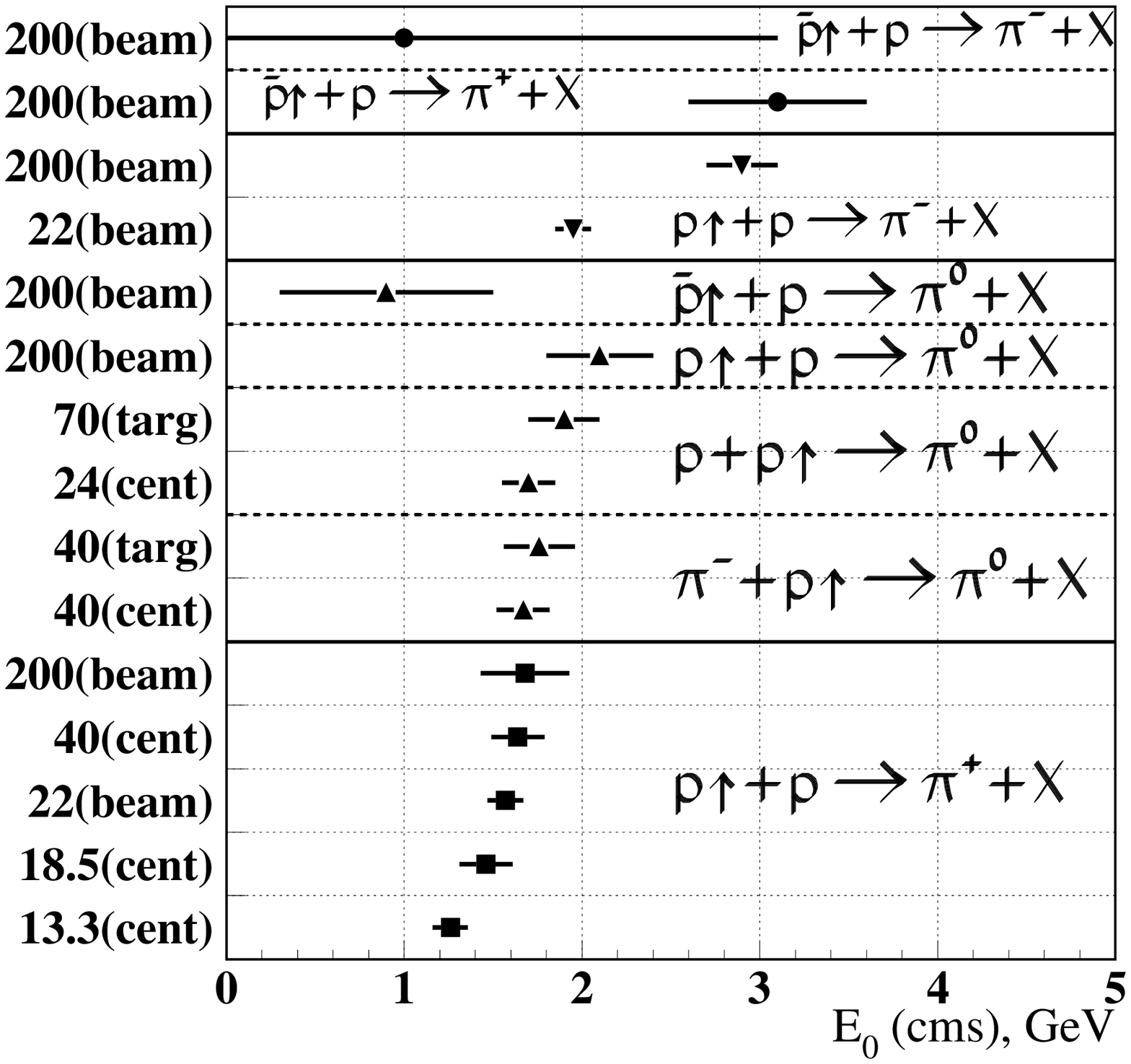,width=9.6cm,height=9.6cm}}
{\small {\bf Figure 3.} Centre of mass energy values where  the pion asymmetry 
begins to grow up for different experiments.
The energy along the Y-axis is in GeV; $cent$ -- corresponds to 
experiments in the central
region ( $x_f\approx0$), $targ$ -- the polarized target
fragmentation region; $beam$ -- the polarized beam fragmentation region. }
\end{wrapfigure}


The comprehensive analysis of all fixed target polarized experiments  
of inclusive $\pi$-meson production was done in  \cite{proza_scaling}.
The result of the analysis is presented in  {\bf Fig.~3}. 

The main conclusion is that the asymmetry begins 
to grow up at the same centre of mass energy $E_0^{cms}=1.5$ to $2.0$~GeV for 
most of the experiments in the energy range between 13 and 200~GeV. 
The analysis was done only for those experimental data where a transverse 
momentum $p_T$ was greater than 0.5~GeV/c to exclude very soft interactions. 
We did not include the experiments when the asymmetry was close to zero.
The conclusion is valid for all $\pi^+$ and $\pi^0$ asymmetries. 
We have to mention that $\pi^-$ production seems to contradict to this. 
We can explain this fact that $\pi^-$-meson at 
small $x_F$ can be produced not only from the valence $d$-quark but also from 
other channels. The interference of different channels is also responsible for 
asymmetry cancellation in $\pi^0$ and $\pi^-$ production in the 
central region. In the reaction $\pimp$ in the central region we found 
significant asymmetry in the contrary to the $\pdupp$ reaction. If in  
the $\pdupp$ reaction the asymmetry is 
cancelled because of different channel interference 
from a polarized and non-polarized proton, in the $\pi^-\pdup$ collisions
the valence $u$-quark from a polarized proton combining with the valence 
$\bar{u}$-quark from $\pi^-$ gives the main contribution to $\pi^0$ 
production, while other channels are suppressed.

In  this scheme the asymmetry behaviour in ${\bar{p}_{\uparrow}p}$ 
interactions in $\pi^+$ and $\pi^-$ production should be inversed in 
comparison with the $\pdupp$ data. The result from E704 
experiment \cite{e704_anti} is consistent with this model. 
The asymmetry of $\pi^+$-production begins to grow up at the same value 
$E_0^{cms} \approx 2.9$~GeV as for $\pi^-$ in reaction $\pdupp$, and 
the asymmetry in the reaction ${\bar{p}_{\uparrow}+p \rar \pi^- +X}$ begins 
to grow up at small value $E^0_{cms}$. 

We can conclude that the meson asymmetry produced by valence 
quark begins to grow up at the same universal energy $E^0_{cms}$.  

The obtained universality of the value $E^0_{cms}$ can manifest
the presence of the universal substructures  in the hadrons --- constituent
quarks. The concept of constituent quark \cite{gmz,mor} has
been  used extensively since the very beginning of the quark era but
has  just obtained recently a  possible
direct experimental evidence at Jefferson Lab \cite{petron}.

A particular model for single spin asymmetries which used the
constituent quark concept in the hadron interaction
was proposed in \cite{spcon}.
The constituent quark appears
as a quasiparticle, i.e. as current valence quark surrounded by
the cloud of quark-antiquark pairs of different flavours, i.e. they are
structured hadron-like objects.
SSA in the model is due to an
 orbital angular momentum of quarks
inside the constituent quark:
spin of constituent quark, e.g. $U$-quark
 is given
by the  sum:
\begin{equation}\label{bal}
J_U=1/2=S_{u_v}+S_{\{\bar q q\}}+L_{\{\bar q q\}}=
1/2+S_{\{\bar q q\}}+L_{\{\bar q q\}}.
\end{equation}
On the grounds of the experimental data for polarized DIS
the conclusion was made  that the significant part of the spin
of constituent quark in the model should be associated with
 the orbital angular momentum
of the current quarks inside the constituent one \cite{spcon}.
In the model SSA reflects internal structure
of the constituent quarks and is proportional to the orbital angular
momentum of current quarks inside the constituent quark.
Evidently, SSA related to the internal orbital momentum
will be non-zero only  when the constituent quark will be excited
and broken up.
The  value $E^0_{cms}$ can be related then to the
minimal energy which is needed for constituent quark excitation and its
dissolution. In this approach it is natural that this energy is universal
since it is adherent to the properties of the constituent quarks. It should be related
anyway to the scale of chiral symmetry breaking $\Lambda_\chi^2$.

Thus the revealed scaling dependence of asymmetry can be interpreted as
 another indication of the presence of constituent quarks in the hadrons.

We are thankful to N.E.~Tyurin for fruitful discussions.

{\it The work is partially supported by Russian Foundation for 
Basic Research grant 03-02-16919}.


\end{document}